\documentclass[aip, apl, a4paper, amsmath,amssymb, reprint]{revtex4-1}

\usepackage{graphicx}
\usepackage{dcolumn}
\usepackage{bm}
\usepackage{hyperref}
\usepackage{units}
\usepackage{upgreek}

\begin{document}

\title[The Strained Silicon Cold Electron Bolometer]{A Strained Silicon Cold Electron Bolometer using Schottky Contacts}

\author{T. L. R. Brien}
\email{tom.brien@astro.cf.ac.uk}
\affiliation{School of Physics \& Astronomy, Cardiff University, Queen's Buildings, The Parade, Cardiff, CF24 3AA, United Kingdom}
\author{P. A. R. Ade}
\affiliation{School of Physics \& Astronomy, Cardiff University, Queen's Buildings, The Parade, Cardiff, CF24 3AA, United Kingdom}
\author{P. S. Barry}
\affiliation{School of Physics \& Astronomy, Cardiff University, Queen's Buildings, The Parade, Cardiff, CF24 3AA, United Kingdom}
\author{C. Dunscombe}
\affiliation{School of Physics \& Astronomy, Cardiff University, Queen's Buildings, The Parade, Cardiff, CF24 3AA, United Kingdom}
\author{D. R. Leadley}
\affiliation{Department of Physics, University of Warwick, Coventry, CV4 7AL, United Kingdom}
\author{D. V. Morozov}
\affiliation{School of Physics \& Astronomy, Cardiff University, Queen's Buildings, The Parade, Cardiff, CF24 3AA, United Kingdom}
\author{M. Myronov}
\affiliation{Department of Physics, University of Warwick, Coventry, CV4 7AL, United Kingdom}
\author{E. H. C. Parker}
\affiliation{Department of Physics, University of Warwick, Coventry, CV4 7AL, United Kingdom}
\author{M. J. Prest}
\affiliation{Department of Physics, University of Warwick, Coventry, CV4 7AL, United Kingdom}
\author{M. Prunnila}
\affiliation{VTT Technical Research Centre of Finland, P.O. Box 1000, FI-02044 VTT Espoo, Finland}
\author{R. V. Sudiwala}
\affiliation{School of Physics \& Astronomy, Cardiff University, Queen's Buildings, The Parade, Cardiff, CF24 3AA, United Kingdom}
\author{T. E. Whall}
\affiliation{Department of Physics, University of Warwick, Coventry, CV4 7AL, United Kingdom}
\author{P. D. Mauskopf}
\affiliation{School of Physics \& Astronomy, Cardiff University, Queen's Buildings, The Parade, Cardiff, CF24 3AA, United Kingdom}
\affiliation{Department of Physics and School of Earth \& Space Exploration, Arizona State University, 650 E. Tyler Mall, Tempe, AZ 85287, United States of America}

\date{\today}

\begin{abstract}
We describe optical characterisation of a Strained Silicon Cold Electron Bolometer (CEB), operating on a $350~\mathrm{mK}$ stage, designed for absorption of millimetre-wave radiation. The silicon Cold Electron Bolometer utilises Schottky contacts between a superconductor and an n\textsuperscript{++} doped silicon island to detect changes in the temperature of the charge carriers in the silicon, due to variations in absorbed radiation. By using strained silicon as the absorber, we decrease the electron-phonon coupling in the device and increase the responsivity to incoming power. The strained silicon absorber is coupled to a planar aluminium twin-slot antenna designed to couple to $160~\mathrm{GHz}$ and that serves as the superconducting contacts. From the measured optical responsivity and spectral response, we calculate a maximum optical efficiency of $50~\%$ for radiation coupled into the device by the planar antenna and an overall noise equivalent power (NEP), referred to absorbed optical power, of $1.1 \times 10^{-16}~\mathrm{\mbox{W\,Hz}^{\nicefrac{-1}{2}}}$ when the detector is observing a $300~\mathrm{K}$ source through a $4~\mathrm{K}$ throughput limiting aperture. Even though this optical system is not optimised we measure a system noise equivalent temperature difference (NETD) of $6~\mathrm{\mbox{mK\,Hz}^{\nicefrac{-1}{2}}}$. We measure the noise of the device using a cross-correlation of time stream data measured simultaneously with two junction field-effect transistor (JFET) amplifiers, with a base correlated noise level of $300~\mathrm{\mbox{pV\,Hz}^{\nicefrac{-1}{2}}}$ and find that the total noise is consistent with a combination of photon noise, current shot noise and electron-phonon thermal noise.
\end{abstract}

\maketitle
\section{Introduction} \label{sec:Introduction}
Photon noise limited detection of millimetre-wave radiation has been demonstrated with a number of cryogenic detectors such as: semiconductor bolometers, transition edge sensors and kinetic inductance detectors\cite{Morozov11,Doyle08}. A bolometer consists of a thermally isolated absorber that converts absorbed radiation into thermal energy, which is detected by means of a sensitive thermometer. The concept of using the weak coupling between electrons and phonons at low temperatures, combined with a normal metal-insulator-superconductor (NIS) tunnel junction thermometer, to make a fast and sensitive hot electron bolometer was first proposed by Nahum, Richards and Mears\cite{Nahum93, Nahum94}. Dual normal metal-insulator-superconductor (SINIS) junctions, coupled to an absorbing metallic island, can be used to simultaneously act as a microrefrigerator by extracting heat from the electrons and as a bolometric detector. The wavelengths that the island absorbs can be defined by patterning the superconducting leads into an antenna.
\begin{figure}[ht]
\includegraphics[width = 0.5\columnwidth]{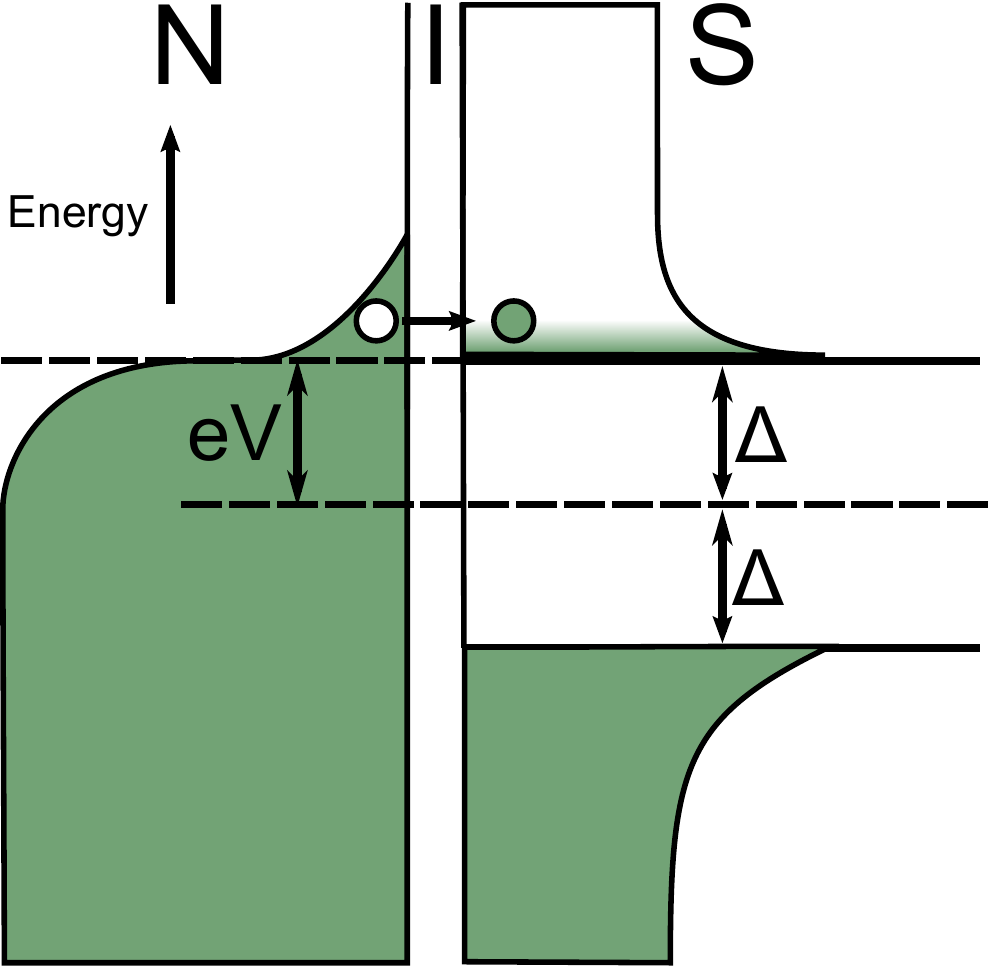}
\caption{Energy bands for a biased normal metal-insulator-superconductor NIS structure. In order for electrons to tunnel from the normal metal (left) into the superconductor (right), we require that  $eV > \Delta - k_{B}T_{e}$ , where $V$  is the voltage across the structure due to the bias, $T_{e}$ is the electron temperature and $\Delta $ is half the superconducting gap.}
\label{fig:NISenergy}
\end{figure}

Schmidt \textit{et al.}\cite{Schmidt2005} describe how the use of a combined microwave and DC biasing signal, along with frequency domain multiplexing techniques, can be used to realise large imaging arrays (up to $10^{5}$ pixels) of cold electron bolometers.

Detailed calculations of the characteristics of these Cold Electron Bolometers (CEBs) indicate that they should exhibit a combination of fast response speeds ($<1~\mathrm{\upmu s}$) and high sensitivity. Achieving high sensitivities with metal-based Cold Electron Bolometers requires fabrication of submicron metal islands. 

Replacing the normal metal with a degenerately doped silicon offers reduced electron-phonon coupling compared to standard metals and thus gives higher sensitivity for a given island volume\cite{Leoni99}. It has been proposed\cite{Muhonen2011} that using a strained silicon absorber enables fabrication of detectors with photon noise limiting sensitivity using standard photolithographic techniques. Initial reports of optical noise equivalent power for metal based Cold Electron Bolometers have been published in recent years\cite{Otto2013, Tarasov2011}. Most of these measurements have been based on radiation absorbed from a cold blackbody source, this does not allow for the spectral response of the detector to be studied. They have also all reported optical noise equivalent powers limited by the readout electronics. Here we present optical measurements of a Strained Silicon Cold Electron Bolometer designed to absorb millimetre-wave radiation, these measurement have been taken with the detector looking out of a window in the cryostat which allowed for a number of sources, including a Fourier transform spectrometer, to be observed.

\section{Theory} \label{sec:Theory} 
The electrothermal properties of both the normal metal-insulator-superconductor and the symmetric (SINIS) structure have been well studied \cite{Pekola05, Nahum93, Nahum94, Leivo96, Savin01, Pekola04}. FIG.~\ref{fig:NISenergy} shows a typical normal metal-insulator-superconductor structure (shown in the presence of an external bias such that $eV = \Delta$). These devices have been shown\cite{Pekola04} to be able to reduce electron temperature from $300~\mathrm{mK}$ to below $100~\mathrm{mK}$. For a sensitive bolometric detector we would like the absorber (the normal metal in this case) to have as small a volume as possible.

A similar structure, superconductor-semiconductor-superconductor (SSmS), exists where the normal metal is replaced by a doped semiconductor and the insulator replaced by a Schottky contact formed between the semiconductor and the superconductor\cite{Savin01}. These devices have the advantage of decreased electron-phonon coupling compared to the normal metal based type of device\cite{Muhonen2011} and reduced electron density. The current, $I$, flowing through each of the symmetric junctions is given by:
\begin{align}
I &=\frac{1}{eR_{N}}\int_{\Delta}^{\infty} \frac{E}{\sqrt{E^{2}-\Delta^{2}}}  \times \left[f\left(E-\nicefrac{eV}{2},T_{e}\right) - \right. \nonumber \\
&\qquad \qquad \qquad \qquad \qquad \qquad  \; \left. f\left(E+\nicefrac{eV}{2},T_{e}\right)\right] \mathrm{d}E\, , \label{eq:IV}
\end{align}
where $R_{N}$ is the normal-state resistance due to tunnelling through the insulating barrier, $\Delta$ is half the superconducting bandgap, $V$ is the  voltage across the structure and  $f\left(E,T\right)$ is the Fermi distribution for electrons at temperature $T_{e}$. Associated with this current is a flow of heat from the central island which dissipates a power, $P$, within the device of:
\begin{align}
P &= IV + \frac{2}{e^{2}R_{N}}\int_{\Delta}^{\infty} \frac{E^{2}}{\sqrt{E^{2}-\Delta^{2}}} \times \left[2f\left(E,T_{s}\right) - \right. \nonumber \\ 
&\qquad \qquad  \left. f\left(E-\nicefrac{eV}{2},T_{e}\right)- f\left(E+\nicefrac{eV}{2}, T_{e}\right)\right] \mathrm{d}E\, \label{eqn:Pc}.
\end{align}
This power is bias dependent and is negative (cooling) for bias voltages $eV \lesssim 3\Delta$.

In a Cold Electron Bolometer when the absorber is heated by incident optical power it is this cooling power, associated with the most energetic of charges tunnelling out of the absorber, which removes the heat. Since the cooling (thermal resetting) of the bolometer is carried out directly by electron diffusion (as opposed to the long, weak, thermal links required by many of today's most sensitive bolometers \cite{Mauskopf97, Audley12, Holland13}) the thermal time constant associated with the Cold Electron Bolometer is governed by the tunnelling time. This can be \cite{Kuzmin04} as low as $10~\mathrm{ns}$, whereas other types of detector \cite{Jackson12} have response times of the order of $1~\mathrm{ms}$.

In addition to this cooling power, the electrons are also heated or cooled by the weak thermal link to the phonons. This heating term, $P_{e-ph}$, is given by:
\begin{align}
P_{e-ph} &= \Sigma \Omega \left(T^{\beta}_{e} - T^{\beta}_{ph} \right), \label{eq:Pe-ph} \\
\intertext{where $\Sigma$ is a material constant that has been measured\cite{Prest11} to be $2 \times 10^{7}~\mathrm{W\,K^{-6}\,m^{-3}}$; $\Omega$ is the volume of the bolometer's absorber; $T_{ph}$ and $T_{e}$ are the phonon and electron temperatures respectively and the power $\beta$ has been found\cite{Prest11} to be $6$. From this we can define a thermal conductance, $G$, from the phonons to the electrons as:}
G = \frac{\mathrm{d}P}{\mathrm{d}T_{e}} &= \beta \Sigma \Omega T^{\beta-1}_{e}.
\end{align}
The total noise equivalent power (NEP) for the Cold Electron Bolometer is comprised of several terms and has been fully derived by Golubev and Kuzmin (2001)\cite{Golubev01} to be:
\begin{align}
NEP^{2}_{CEB} &= \frac{\left<\delta V^{2}\right>_{\mathrm{amp}}}{S^{2}} + 2\beta k_{B} \Sigma \Omega\left(T_{e}^{\beta+1} + T_{ph}^{\beta+1}\right) \nonumber \\
&\qquad+ \left<\delta P^{2} \right> - 2\frac{\left<\delta P \, \delta I\right>}{\nicefrac{\partial I}{\partial V}S} +\frac{\left<\delta I^{2}\right>}{\left(\nicefrac{\partial I}{\partial V}S\right)^{2}} \, , \label{eq:CEB_NEP}
\end{align}
where $\left<\delta V^{2}\right>_{\mathrm{amp}}$ is the noise of the readout amplifier and $S$ is the responsivity of the detector, which is a function of bias, $\left<\delta P\right>$ is the heat flow noise and $\left<\delta I\right>$ is the current noise. The use of strained silicon reduces the constant $\Sigma$ by a factor of $25$ compared to unstrained silicon\cite{Prest11}, this results in a corresponding improvement in the second term of EQN.~\ref{eq:CEB_NEP} (the phonon noise).

The other dominant limiting factor to the noise equivalent power will be due to the absorption of photons into the strained silicon. This photon noise term is:
\begin{align}
NEP^{2}_{photon} &= 2h\nu P_{opt} + \frac{P_{opt}^{2}}{\delta \nu}, \label{eq:photonNEP}
\end{align}
where $\nu$ and $P_{opt}$ are the frequency and power of the incident radiation respectively and $\delta \nu$ is the optical bandwidth.
\par

\section{Device Design} \label{sec:Device}

\begin{figure}[ht]
\includegraphics[width = 0.8\columnwidth]{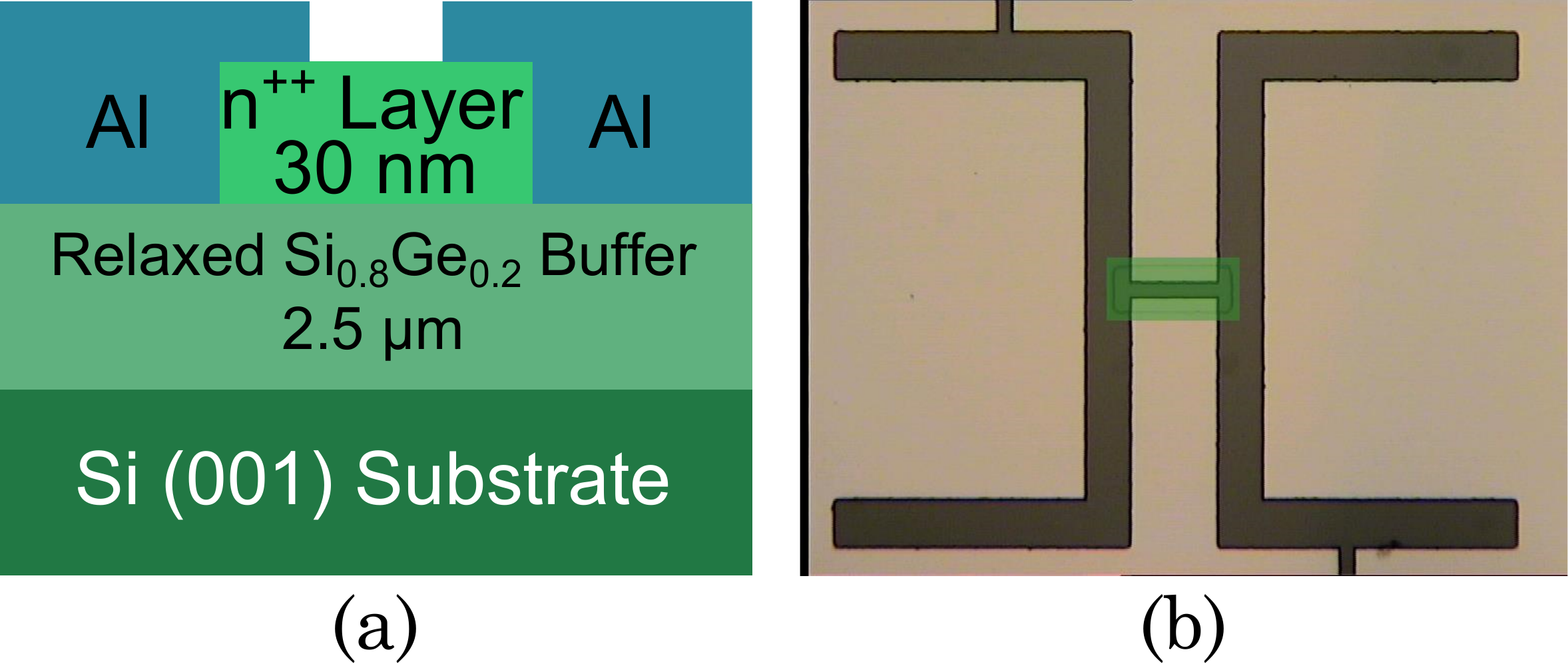}
\caption{(a) Cross-sectional view of Cold Electron Bolometer structure. (b) Optical image a Cold Electron Bolometer. A small island absorber of n\textsuperscript{++} doped silicon ((a) - green, (b) - highlighted green) sits atop a strained SiGe virtual substrate ((a) - light green, (b) - brown); the top layer of aluminium ((a) - blue, (b) - beige) forms both the antenna structure and the contacts to the absorber; the small slots, which can be seen at the edges of the device, allows DC measure of the cold electron bolometer without affecting the antenna coupling.}\label{fig:device_design}
\end{figure}

One advantage of the silicon based Cold Electron Bolometer compared to those utilising a metal absorber (SINIS) is that since the tunnel barrier is formed by a Schottky contact the is no need to fabricate separate insulating layers. The Strained Silicon Cold Electron Bolometer, studied in this work, consists of three elements: Firstly, the silicon substrate has an epitaxially grown $2.5~\mathrm{\upmu m}$ thick relaxed SiGe (80 \% silicon) straining layer.  On top of the straining layer is a $30~\mathrm{nm}$ thick layer of n\textsuperscript{++} doped silicon ($N_{D} = 4 \times 10^{19}~\mathrm{cm}^{-3}$) etched to form a rectangular mesa with an area of $38~\mathrm{\upmu m} \times 14~\mathrm{\upmu m}$. Finally the top layer is a $100~\mathrm{nm}$ thick film of e-beam evaporated aluminium. This final layer is patterned to form both the contacts to the doped silicon absorber and a twin slot antenna. The contacts to the absorber are both $30~\mathrm{\upmu m} \times 5~\mathrm{\upmu m}$ and have a give a tunnelling resistance of $290~\mathrm{\Omega}$. The twin slot antenna has been designed to couple $160~\mathrm{GHz}$ radiation to the central absorber and the coupling has been simulated with Ansoft's HFSS software prior to fabrication. The device design is shown in FIG.~\ref{fig:device_design}.

\section{Experimental Procedure}\label{sec:exp_procedure}
\begin{figure}[ht]
\includegraphics[width = 0.8\columnwidth]{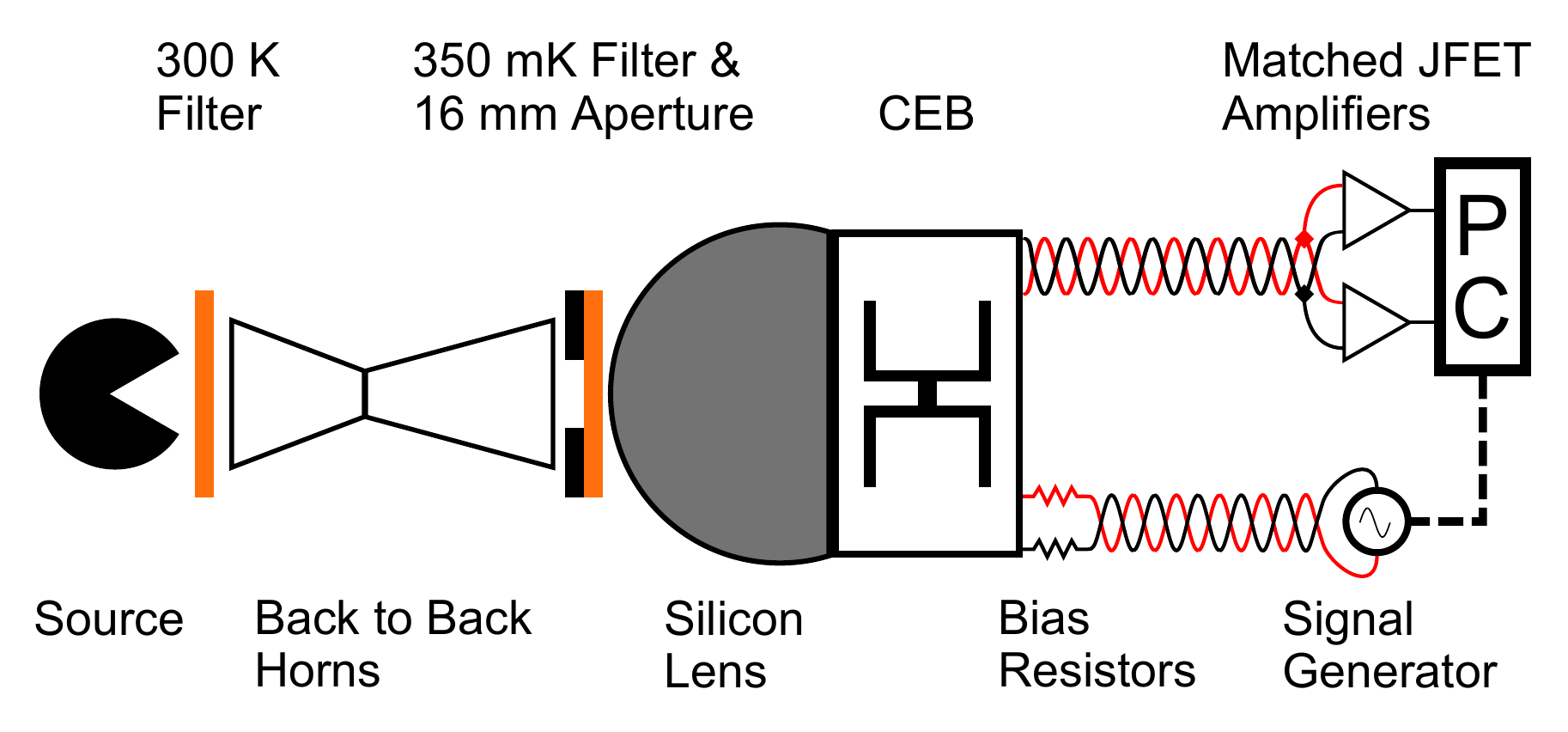}
\caption{Experimental setup, radiation is focussed onto the detector chip via a pair of back-to-back horns and a silicon lens. Optical filters are placed before and after the horns and have the effect of limiting the radiation seen by the detector to an upper limit of $300~\mathrm{GHz}$. The detector is biased via a simple voltage generator and biasing resistors. The voltage output of the detector is sent into two JFET based amplifiers (each with an input referred noise of $2~\mathrm{\mbox{nV\,Hz}^{\nicefrac{-1}{2}}}$) and the output of these is correlated to achieve a final input referred noise of $300~\mathrm{\mbox{pV\,Hz}^{\nicefrac{-1}{2}}}$.}
\label{fig:exp_setup}
\end{figure}

A schematic of the testing setup is shown in FIG. \ref{fig:exp_setup}. The detector was housed in a liquid helium cryostat and cooled to $350~\mathrm{mK}$ using a helium-3 refrigerator. Radiation, visible through a window in the outer cryostat shield, was fed in to a pair of back-to-back horns, the beam from this horn pair was then focussed on to the detector's antenna by a hemispherical silicon lens. This optical coupling scheme was not optimised for high efficiency but designed to minimise stray light coupling to the device.

The detector was current biased using a differential voltage source and a pair of cold $1~\mathrm{M\Omega}$ biasing resistors. The voltage output of the detector was fed into two matched JFET differential amplifiers, each of which had an input referred noise of $2~\mathrm{\mbox{nV\,Hz}^{\nicefrac{-1}{2}}}$. The output of each of these amplifiers was then passed to a computer which cross-correlated the signal in real time and resulted in a final input referred correlated noise, after averaging, of $300~\mathrm{\mbox{pV\,Hz}^{\nicefrac{-1}{2}}}$ for the readout system. For optical testing we used an eccosorb load chopped between $ 300~\mathrm{K}$ and $77~\mathrm{K}$.

\section{Results} \label{sec:Results}
\begin{figure}[ht]
\includegraphics[width = 0.8\columnwidth]{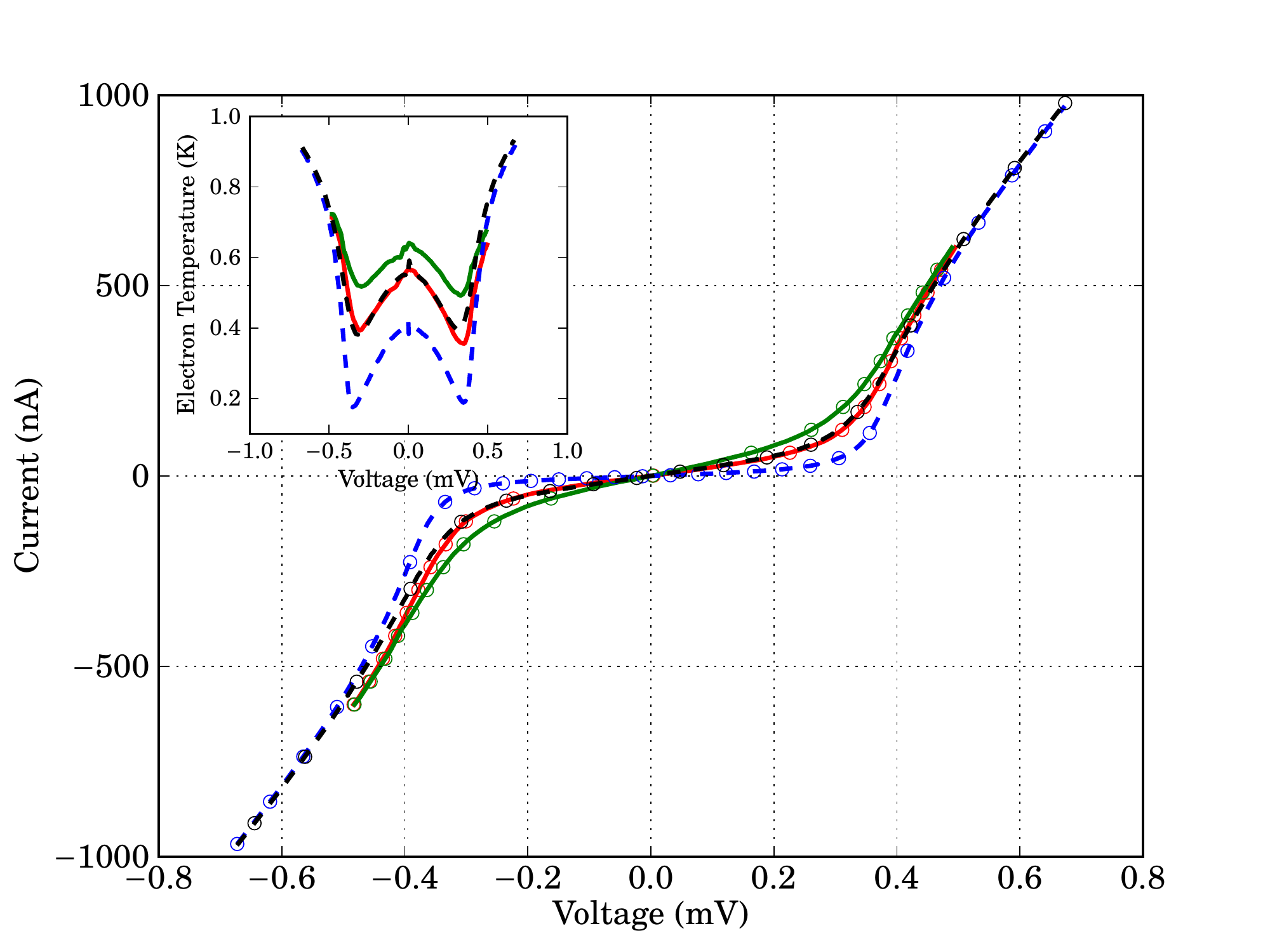}
\caption{IV characteristics and model fit. Solid lines optical measurements; dashed lines dark measurements. Red - $77~\mathrm{K}$ source; Green - $300~\mathrm{K}$ source; Blue - $T_{ph} = 350~\mathrm{mK}$; Black - $T_{ph} = 550~\mathrm{mK}$ . There is a clear shift of the IV towards the linear as the incident power is increased. Lines - Model fit based on $T_{e}$ fitting in EQN. \ref{eq:IV}. Circles - Heavily reduced experimental data. Inset - Variation in electron temperature with bias; colours as in main figure.}
\label{fig:IV_data_model}
\end{figure}
The Silicon Cold Electron Bolometer has been tested both dark and optically loaded. Dark measurements consist of current-voltage (IV) characterisation at various bath (phonon) temperatures. The optical response of the device to a variable temperature blackbody source has also been measured. FIG.~\ref{fig:IV_data_model} compares the current-voltage relationship for the detector in these various conditions; it can be seen that the optically loaded measurements correspond to higher electron temperature in the device and therefore more linear current-voltage curves compared to the corresponding unloaded measurement. In fact the optically loaded curves are similar to a dark measurement at a much higher phonon temperature.

From the measured voltage for a given current bias and using EQN.~\ref{eq:IV} we can calculate the temperature of the electrons. This model, shown as the lines in FIG.~\ref{fig:IV_data_model}, shows that a high quality fit to the data (open circles) can be achieved based on this algorithm in all cases. The electron temperatures found from this fit were $570~\mathrm{mK}$ and $640~\mathrm{mK}$ at zero bias for the $77~\mathrm{K}$ and $300~\mathrm{K}$ illuminations. The increase from the phonon temperature of $350~\mathrm{mK}$ is accounted for by the incident power heating the electrons. At a bias corresponding to a voltage of $\sim 2\Delta$ across the detector, the minimum electron temperatures achieved for the two illumination levels were $350~\mathrm{mK}$ and $500~\mathrm{mK}$. By use of EQN.~\ref{eq:Pe-ph} at zero bias, combined with the dimensions of the absorbing island and the measured value of $\Sigma$ $(2.7 \times 10^{7}~\mathrm{W\,K^{-6}\,m^{-3}})$ and assuming the electron temperature is significantly greater than that of the phonons, we compute the absorbed power to be $10.5~\mathrm{pW}$ \& $21.5~\mathrm{pW}$ for the two load temperatures. We believe there is a contribution of approximately $5~\mathrm{pW}$ from stray light to both of these powers.
\begin{figure}[t]
\includegraphics[width = 0.8\columnwidth]{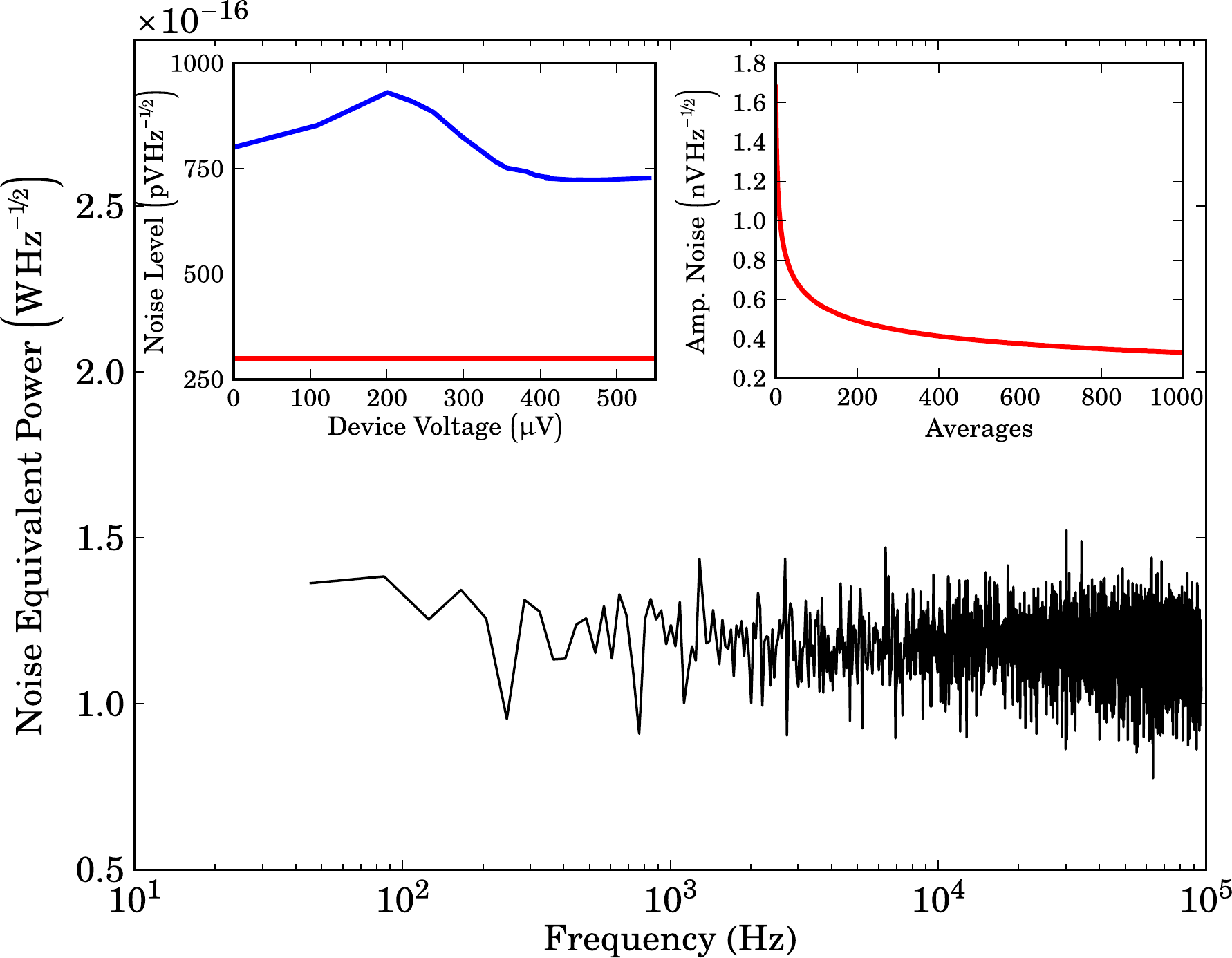}
\caption{Noise equivalent power for a SiCEB, as a function of readout frequency, operating at optimum bias ($eV=2\Delta$) with $10.5~\mathrm{pW}$ of absorbed optical power. Left inset - Measured device noise (blue) and amplifier noise limit (red). Right inset - Reduction in amplifier noise with averaging for two JFET amplifiers operating in cross-correlated mode.}
\label{fig:NEP_ampNoise}
\end{figure}

The responsivity, at a particular current bias, of the Cold Electron Bolometer can be calculated from the change in the voltage when the incident power changes by a known amount. From the calculated absorbed powers for the two illuminations and the voltage changes resulting from this change (seen in  FIG.~\ref{fig:IV_data_model}), we calculate the responsivity to have a maximum of $7.9 \times 10^{6}~\mathrm{\mbox{V\,W}^{-1}}$ for the $77~\mathrm{K}$ $(10.5~\mathrm{pW})$ source and $2.8 \times 10^{6}~\mathrm{\mbox{V\,W}^{-1}}$ for the room temperature $(21.5\mathrm{pW})$ source. In both cases the maximum responsivity occurs when the voltage across the device is just below $2\Delta$, as is expected. FIG.~\ref{fig:NEP_ampNoise} shows the noise equivalent power calculated from these results. For both the $77\mathrm{K}$ \& the $300~\mathrm{K}$ loading this is dominated by photon noise. From FIG.~\ref{fig:NEP_ampNoise} we see that the $77~\mathrm{K}$ noise equivalent power is $1.1 \times 10^{-16}~\mathrm{\mbox{W\,Hz}^{\nicefrac{-1}{2}}}$.

The speed of the detector can be found from the roll-off in the white noise level from the photon noise or from measuring the change in responsivity for a modulated signal as a function of frequency. We attempted to measure this using a coherent $150~\mathrm{GHz}$ tunable source which could be chopped on and off at frequencies up to 6~kHz but did not see any reduction in the signal and we also did not see any roll-off in the noise power (as seen in FIG.~\ref{fig:NEP_ampNoise}) up to the bandwidth of the readout amplifier $\left(100~\mathrm{kHz}\right)$. From this, we conclude that the time-constant of this detector is less than $1~\mathrm{\upmu s}$.

From EQN.~\ref{eq:CEB_NEP} we compute that the limit on the electrical (dark) noise equivalent power, for optical loading less than $1\mathrm{pW}$, from the electron-phonon interaction is $8.3 \times 10^{-18}~\mathrm{\mbox{W\,Hz}^{\nicefrac{-1}{2}}}$, this compares well to the `dark' noise equivalent power estimations for hot electron bolometer type devices operating at comparable phonon temperatures\cite{Karasik2011}, which share a common noise limit in these circumstances. The current proof of concept detector has a very large absorbing element, if this was reduced by a factor of $10$ (which is still larger than the absorbing element of the comparable hot electron bolometer\cite{Karasik2011} and still possible with standard photolithography) the phonon noise limit would be reduced to $2.6\times 10^{-18}~\mathrm{\mbox{W\,Hz}^{\nicefrac{-1}{2}}}$ for the same operating temperature.
\begin{figure}[ht]
\includegraphics[width = 0.8\columnwidth]{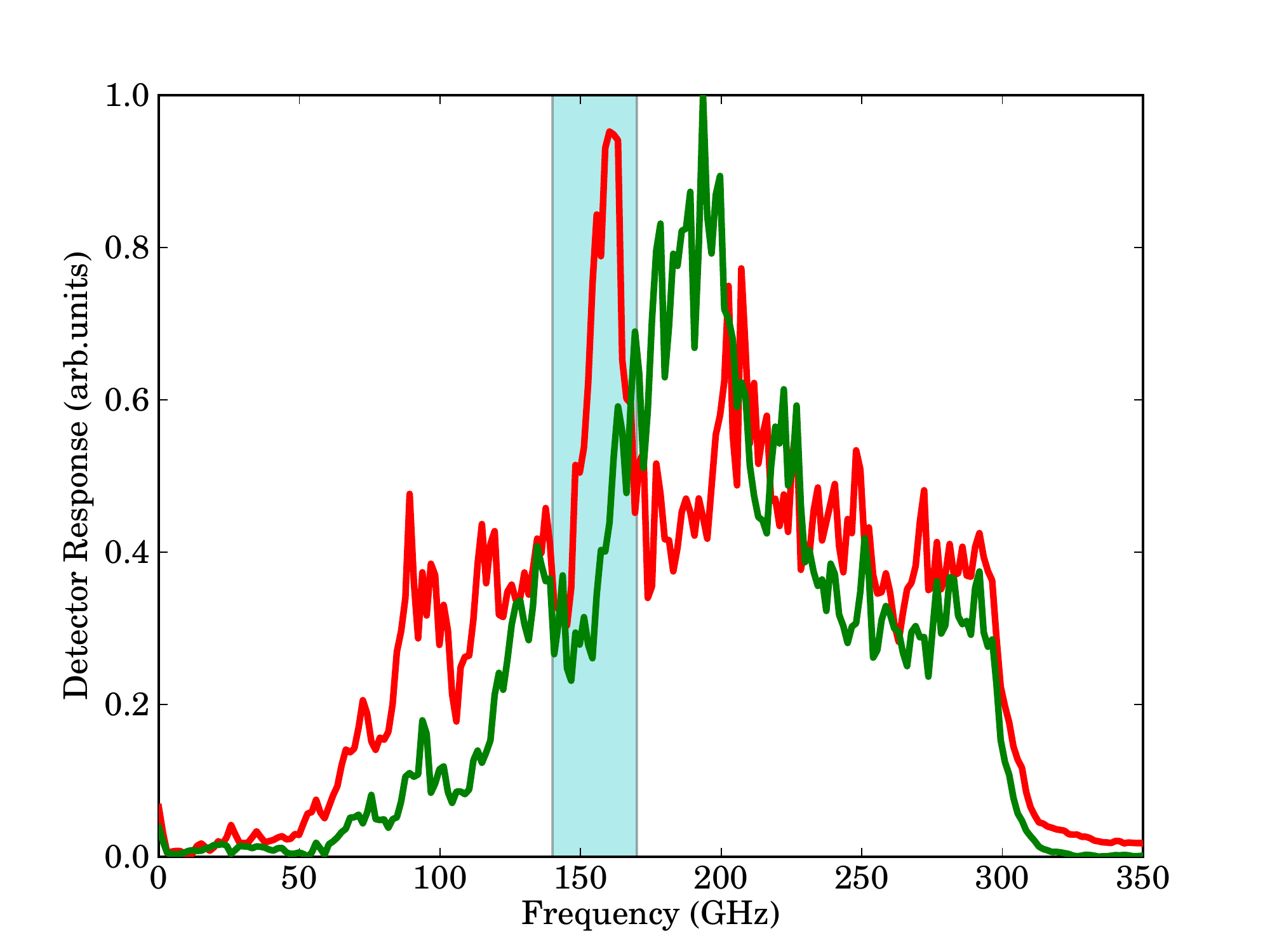}
\caption{Response of the Strained Silicon Cold Electron Bolometer to a Fourier Transform Spectrometer with a mercury arc lamp source. Red - Response to a vertically polarised source; green - Horizontally polarised source; highlighted region - expected frequency range of the antenna's $3~\mathrm{dB}$ response.}
\label{fig:FTS_response}
\end{figure}

We have also measured the response of the Strained Silicon Cold Electron Bolometer as a function of the frequency of incident radiation. This was performed in both linear polarisations; since the detector used a twin-slot antenna to couple radiation it was expected that there would be more response in one polarisation. The measured spectral response is shown in FIG.~\ref{fig:FTS_response}. The measured response has a cutoff of $300~\mathrm{GHz}$ due to the optical filters in place. The highlighted region denotes the expected frequency range of the twin-slot antenna. There is a clear excess response in this region in the vertical polarisation, parallel to the twin slot antenna. The peak in the horizontal polarisation may be attributed to response in the coplanar waveguide (CPW), which couples radiation to the absorber and is also due to the cuts in the aluminium (seen in FIG.~\ref{fig:device_design}b), which break the DC continuity around the detector. Both these cuts and the coplanar waveguide are orthogonal to the twin-slot antenna. The plateau level, around half of the maximum response, is due to a combination of photons directly splitting Cooper pairs in the aluminium along with direct absorption in the doped silicon mesa, general broadening of the absorption spectrum due to the silicon lens and the integrating cavity in which the detector was housed.

\section{Conclusion}\label{sec:conclusion}
We have demonstrated a detector that utilises direct electron cooling via Schottky tunnelling contacts between aluminium and strained silicon. We have shown that this detector has a photon noise limited noise equivalent power of $1.1 \times 10^{-16}~\mathrm{\mbox{W\,Hz}^{\nicefrac{-1}{2}}}$ when observing a  $77~\mathrm{K}$ blackbody and under low optical loading conditions has an electrical or dark noise equivalent power, at $350~\mathrm{mK}$, of $8.3 \times 10^{-18}~\mathrm{\mbox{W\,Hz}^{\nicefrac{-1}{2}}}$. The time constant of this detector has been determined to be less than $1~\mathrm{\upmu s}$, which compares extremely favourably to other detector types with similar noise equivalent power.

This work has been financially supported by the EPSRC through grant numbers EP/F040784/1 and EP/J001074/1, and by the Academy of Finland through grant number 252598.
\bibliography{bib}
\end{document}